# First highly pathogenic avian influenza outbreak in a commercial farm in Brazil: outbreak timeline, control actions, risk analysis, and transmission modeling

Nicolas C. Cardenas[1], Francisco N.P. Lopes[2], Paulo A. S. C. de Souza[2], Fernando H.S. Groff[2], Ananda P. Kowalski[2], Alessandra Krein[2], Rodrigo N. Etges[2], Daniela L. de Azevedo[2], Alencar Machado[3], Vinícius Maran[3], Felipe A. Machado[3], Gustavo Machado[1*]

[1]Department of Population Health and Pathobiology, College of Veterinary Medicine, North Carolina State University, Raleigh, NC, USA.

[2]Departamento de Defesa Agropecuária, Secretaria da Agricultura, Pecuária e Desenvolvimento Rural, Porto Alegre, Brazil.

[3] Laboratory of Ubiquitous, Mobile and Applied Computing (LUMAC), Polytechnic College of Federal University of Santa Maria, Santa Maria, Brazil.

*Corresponding author: gmachad@ncsu.edu.

**Abstract**

On May 15, 2025, Brazil reported its first highly pathogenic avian influenza (HPAI) outbreak in a commercial poultry breeder farm in Montenegro, Rio Grande do Sul. This study presents the outbreak timeline, control measures, along with spatial risk assessment and epidemiological model used to simulate detection delays. The transmission model considered Susceptible-Exposed-Infected-Recovered/Dead farm statuses to simulate within-farm and between-farm dynamics under 3-, 5-, and 10-day detection delays. The single infected commercial farm lost 15,650 birds, with 92% mortality due to HPAI, and additional culling of the remaining birds on Day 5 post-notification to the state animal health officials. Based on the mortality and outbreak response data, the introduction likely occurred 3–10 days before its official detection. Our field investigations suggested that wild birds were the most likely source of introduction, although biosecurity breaches could not be ruled out. Control measures implemented included movement restrictions and a control zone, from which 4,197 vehicles were inspected upon entry. Risk analysis classified 64.4% of municipalities as low risk, 35.0% as medium risk, and 0.6% as high risk. Our HPAI disease simulation results showed that the number of secondary infections would increase from a median of 4 farms (IQR 2–5) with a 3-day delay to 6 (IQR 3–22) and 34 (IQR 12–47) farms with 5- and 10-day delays, respectively. The rapid veterinary response eliminated the outbreak within 32 days of detection, highlighting the critical role of early detection and prompt response.

## 1. Introduction

Avian influenza virus (AIV) is a zoonotic pathogen that affects a wide range of species, including wild birds and domestic birds (Kirkeby and Ward, 2022). AIVs pose significant challenges due to their global distribution and high mortality in birds, often decimating entire commercial flocks within days (Charostad et al., 2023; Goodman et al., 2025). Wild birds, such as *Anas platyrhynchos*, serve as natural reservoirs of AIVs, while migratory species often exhibit higher prevalence and act as primary vectors (Capua and Alexander, 2007). These birds are frequently asymptomatic and are resistant to highly pathogenic avian influenza (HPAI) (McDuie et al., 2022). Through flyway corridors such as the Atlantic and Mississippi routes, they can carry AIVs for long distances, connecting bird populations from North America, the Caribbean, and Europe to wintering or stopover sites in South America (Chesser, 1994).

      Given the recent large-scale outbreaks in the United States, which have resulted in 325 cases on commercial poultry farms (USDA, 2023), and the Atlantic flyway, the changes in HPAI between the American continent pose a significant risk to the Brazilian poultry industry. The introduction of HPAI in Brazil, the world's largest exporter of chicken meat, is a significant concern, as it accounts for approximately 35–38% of international trade, reaching a record 5.29 million tons valued at nearly US$10 billion. Brazil supplies poultry products to around 150 to 160 countries worldwide (ABPA, 2024).

      Since 2023, Brazil has notified the World Organisation for Animal Health (WOAH) of 148 HPAI cases in wild birds across seven states, including Espírito Santo, Rio de Janeiro, São Paulo, Bahia, Paraná, Santa Catarina, and Rio Grande do Sul (WOAH, 2024). Affected species ranged from seabirds and aquatic birds to raptors, the latter of which were likely infected through scavenging of contaminated carcasses. Most detections occurred along beaches and islands frequented by migratory seabirds, with notable die-offs reported in Espírito Santo and Rio de Janeiro. Since 2023, HPAI has also spread to southern Brazil, with 20 cases reported in wildlife (MAPA, 2025). Until 2025, Brazil reported cases of HPAI in a backyard flock in the state of

Espírito Santo. The non-commercial affected flock raised chickens and ducks, which died and tested positive for HPAI. Two additional backyard cases were reported in the state of Mato Grosso. Additionally, HPAI was also detected in a *Thalasseus acuflavidus* specimen found in the municipality of Marataízes, Espírito Santo state (Reischak et al., 2023). An additional three cases were identified in backyard poultry and wild birds; however, no infections were reported in humans or commercial poultry flocks (Reischak et al., 2023).

Mathematical modeling and spatial risk analysis have been increasingly applied to enhance disease spread preparedness by identifying key variables associated with outbreaks and evaluating control strategies (Dembek et al., 2018). These methods provide a framework for assessing surveillance capacity and designing effective interventions before, during, and after the outbreak event (Dembek et al., 2018). Here, we employed a mathematical model and spatial risk analysis to examine the first confirmed outbreak of HPAI in Brazil, which occurred on May 15, 2025, at a commercial breeding layer farm in the municipality of Montenegro, Rio Grande do Sul state (MAPA, 2025). Here, we also described the outbreak timeline and the control measures implemented, including activities carried out by the official veterinary service, from the initial report until the elimination of the outbreak.

## 2. Materials and methods

*Outbreak description*

To reconstruct the outbreak timeline, we systematically compiled and cross-validated information from multiple official sources in collaboration with the Departamento de Vigilância e Defesa Sanitária Animal, Secretaria da Agricultura, Pecuária, Produção Sustentável e Irrigação (SEAPI), Porto Alegre, Brazil (SEAPI, 2024). Primary data were extracted from reports issued by SEAPI and the Brazilian Ministry of Agriculture, Livestock, and Supply (MAPA). A series of official reports titled Centro de Operações de Emergência Zoossanitária Rio Grande do Sul (COEZOO-RS) (Numbers 001/2025 to 023/2025) was used to reconstruct the chronology of field activities and laboratory confirmation. These documents provided detailed information about the first mortality notifications, farm inspections, sample collections, outbreak confirmation, and the implementation of control and elimination measures. Diagnostic reports from the Official Federal Agricultural Defense Laboratory (LFDA) in Campinas, São Paulo, Brazil. Each entry in the timeline represents a verified event, with dates cross-checked against multiple official records provided by SEAPI's Official Veterinary Service (OVS). The complete timeline is presented in Supplementary Material Table S1. The main outbreak events included: i) the implementation of control the area; ii) road barriers and the daily number of vehicles cleaned and disinfected; iii) the daily number of staff directly involved with stamping out the outbreak. Data used in forecasting HPAI spread and spatial risk analysis included the geospatial locations of all poultry farms (both commercial and backyard farms), farm populations, production types, and farm-to-farm movement data for commercial poultry farms.

## 2.1. Datasets

*Population and movement data*

The population data was provided by the Agricultural Defense System (SDA), which encompasses 95,276 poultry farms across Rio Grande do Sul (SDA, 2024). Of those records, 6,815 (7.15%) had inconsistencies and were not used. The final dataset includes 89,349 valid records. The vast majority, 84,776 (94.88%) of the total, are backyard farms, accounting for 14,638,399 birds. From 4,573 commercial farms with population data, the total number of birds was 215,587,378, with geolocation and recorded population sizes (Figure 1). Farm-to-farm commercial birds and eggs movement data provided by SDA included 50,129 records between March 1, 2025, and May 14, 2025.

*Outbreak data*

During the outbreak response, daily operational data were collected, which included the number of farms visited for surveillance, the number of staff, and the number of vehicles crossing for cleaning and disinfection at road barriers. The outbreak response activities were recorded using the "Plataforma de Defesa Sanitária Animal do Rio Grande do Sul" (PDSA) from May 18, 2025, to June 17, 2025 (Machado et al., 2025; PDSA-RS, 2025). Briefly, the PDSA is a real-time digital management tool used by SEAPI to operationalize several livestock programs, encompassing the entire process from the producer's request for certification through laboratory testing and certification issued by the federal government. The full description of the system is provided in the Supplementary Material Section of the PDSA-RS Platform. PDSA has one module that enables SEAPI to operationalize outbreak response by (i) creating control zones and (ii) assigning field staff for surveillance and farm visits, (iii) as well as the dynamic creation of data collection checklists, and (iv) animal and eggs traceability, which are fed back to PDSA and managed by the SEAPI incident command. Here, we used the data collected by the checklist that was completed at every farm visit in the infected, buffer, and surveillance zones.

*Environmental spatial data*

To analyze the spatial risk distribution of HPAI, we retrieved a series of raster data at the national level, which included: mean surface temperature [WorldClim v2.1] (Wan et al., 2025), Normalized Difference Vegetation Index (NDVI)[MODIS/006/MOD13A1] (Didan, 2025), and water occurrence frequency [JRC/GSW1_3] (Brown et al., 2022), which were generated at 1 km resolution for April–May 2025 via Google Earth Engine (https://developers.google.com/earth-engine/datasets/). Vector layers comprised municipality boundaries of Brazil (Prado Siqueira, 2021) and wild bird data, providing area-based estimates of wild bird abundance in Brazilian territory, retrieved from ICMBio (2025).

*2.2 Mechanistic modeling, forecasting the outbreak dynamics*

A mechanistic, single-pathogen model using a Susceptible-Exposed-Infected-Recovered/Dead was implemented via our Multi-Host Animal Spread Stochastic Multilevel Model
The MHASpread framework (Cardenas et al., 2025; Cespedes Cardenas and Machado, 2024) and used to simulate the HPAI outbreak. The HPAI model incorporated within-farm dynamics, taking into account bird-to-bird transmission and between-farm transmission through the movement of birds, as well as spatial transmission (Supplementary Material Methods section and Supplementary Material Table S2). Briefly, a kernel transmission approach is employed in spatial transmission, in which the transmission probability decreases as the distance between farms increases. The birds' movements are modeled to replicate their actual movements, taking into account the health status of the birds at both the origin and destination.

We ran 1,000 simulations in which the initial outbreak started from the detected farm in Montenegro. We varied the introduction date by simulating delays of three, five, and 10 days in the detection, given the uncertainty of the exact introduction date of this first HPAI case in a commercial poultry farm in Brazil. We used the MHASpread model to estimate the daily cumulative probability of secondary infection and guide the outbreak response. The cumulative probability of outbreaks was transformed into an empirical cumulative distribution function (eCDF) by converting the number of infected farms (Ayer et al., 1955). Results were subsequently aggregated within 10 km hexagonal spatial grids and used to generate thematic maps of infection probability, which were shared with animal health officials during the first days of the outbreak response.

## 2.3 Spatial risk analysis

First, we evaluated the spatial distribution of HPAI cases in Brazil, thus incorporating previously reported cases in backyard flocks and wild birds at the national level. To determine whether HPAI cases were clustered or randomly distributed across the Brazilian territory, we applied Moran's I test (Moran, 1950). We then fitted spatial models that incorporated environmental covariates and HPAI outbreak data. Spearman correlation coefficients were calculated, and thematic maps were produced to explore associations among variables. The spatial autocorrelation was assessed using Moran's I statistic, and spatial weight matrices were generated based on queen contiguity neighborhoods at the national level in Brazil. We evaluated four distinct statistical and machine learning approaches. Generalized Linear Model (GLM) with Poisson Distribution, Spatial Lag Model (SAR) (Bidanset and Lombard, 2014), Spatial Error Model (SEM) (Yildirim and Mert Kantar, 2020), incorporating spatial weights to account for geographic dependencies in the data, and a Random Forest model (Liaw and Wiener, 2002). For each spatial analysis, we implemented a variable selection process as follows: All candidate variables (temperature, water occurrence, NDVI, and migratory bird concentration) were included in the models. Variable importance and selection were assessed using model-specific metrics (p-values screening or variable importance scores for Random Forest). Variables with non-significant contributions or negative importance scores were systematically removed. The most parsimonious model with optimal performance was evaluated and retained using multiple metrics appropriate for both regression and imbalanced classification, as calculated by the root-mean-square error (RMSE) and the mean absolute error (MAE) (Chai and Draxler, 2014). Of note, before model fitting, multicollinearity among predictor variables was assessed and addressed through correlation analysis. No strong correlation was observed in the selected variables. The final fitted model, based on the RMSE and MAE, was the spatial autoregressive model (SAR), as described in detail in Equation 1.

$$Risk_i = \rho W \cdot Outbreaks + \beta_0 + \beta_1 Migratingbirds + \beta_2 Wateroccurrence + \varepsilon \quad (1)$$

Where $\rho$ represents the spatial autoregressive parameter, $W$ is the spatial weights matrix, and $\varepsilon$ represents the error term. $Outbreaks$ represents the count of HPAI present at the municipality $i$, reported in Brazil from January 2023 to May 2025 (WOAH, 2024), while $Migratingbirds$ and $Wateroccurrence$ represents the average value of the area of migratory birds reported at the municipality, and the probability of water occurrence, respectively. The risk $Poultryrisk$, considering the commercial poultry farm population, was calculated as follows in Equation 2.

$$Poultryrisk = Risk_i * Normalizedpopulation_i \quad (2)$$

Where $Poultryrisk$ are the fitted values from the SAR model at the municipality $i$, while $Normalizedpopulation$ represents the normalized number of birds in the area using min-max scaling at the municipality $i$, resulting in a thematic map.

**2.4 Software**

This study was conducted using R statistical software version 4.1.1 (R Core Team, 2025) and the following packages: tidyverse (Wickham et al., 2019), raster (J. Hijmans, 2021), MHASpread (Cespedes Cardenas and Machado, 2024), sf (Pebesma, 2018), and Random Forest (Liaw and Wiener, 2002).

## 3. Results

### *3.1. Outbreak description*

A commercial breeding layer farm, in its 35th week of production, exhibited high mortality in barn one, with 8,550 birds, of which 100% died within 72 hours (Figure 2-A). On May 9, birds were found recumbent, displaying loss of motor function characterized by prostration with the neck extended forward, ruffled feathers, tremors, and rigidity of the legs and wings (Figure 2-B-C). When prompted to rise, birds exhibited tachypnea, and some presented with diarrhea. The high mortality was reported to state animal health officials on May 11, 2025. On the same day, barn two (8,458 layers) was inspected, and birds had no clinical signs or mortality (Supplementary Material, Figure S1).

The progression of mortality in each barn is depicted in the Supplementary Material (Figure S1). Notification of a suspected case of Avian Respiratory and Nervous Syndrome (ARN) was then reported to the Official Veterinary Service (OVS) on the night of May 11, 2025 (Day 1), by the veterinarian in charge of the breeding farm. Samples were collected on May 12 (Day 2) and submitted to the official LFDA laboratory on May 13 (Day 3). On 14 May (Day 4), HPAI was confirmed in wild birds at the Sapucaia do Sul Zoo, 26 km from the commercial breeding farm. On May 15 (Day 5), the occurrence of HPAI H5N1 (clade 2.3.4.4b) at the investigated farm was confirmed, leading to the official declaration of an outbreak and communication to the World Organisation for Animal Health (WOAH) regarding event identification 6484. Coordinated by the Centro de Operações de Emergência Zoossanitária (COEZOO-RS), the emergency response involved federal, state, and municipal agencies, as well as support from the military and private sector. At Day 5, the mortality in the zoo had affected 65 birds, including Black Swans (*Cygnus*

*atratus*) and Black-necked Swans (*Cygnus melancoryphus*). The mortality at the zoo is presented in Supplementary Material Figure S2. As a precautionary measure, the zoo was immediately closed to the public to mitigate the risk of further intra- and interspecies transmission. On May 15 (Day 5), the control areas were established around the infected farm (Figure 7), and animal movements and products were blocked in and out of the control zones.

On May 16 (Day 6), the federal government declared a 60-day state of emergency for animal health affecting avian species, through MAPA Ordinance Number 795. This legal step released federal funds and granted regulatory powers to respond to the outbreak, enabling the rapid mobilization of personnel, procurement of disinfectants and equipment, and enforcement of emergency measures, as well as a standstill on the movement of animals and fertilized eggs. Following the confirmation of HPAI, the index farm was placed under strict quarantine. A total of 17,008 layers were affected, of which 15,650 had died from the disease. The remaining 1,358 live birds (8%) of barn two were culled on-site via manual cervical dislocation. Additionally, 70,000 fertile eggs stored on the farm were destroyed. Cleaning and disinfection of the farm, including equipment, barns, silos, and vehicles, as well as disposal of birds and eggs under SEAPI and MAPA supervision following biosafety protocols, began on Day 6 and was completed on Day 11. In addition, epidemiological investigations of the infected farm were conducted to trace potential sources of virus introduction, including feed truck movements, fertile egg trucks, human activity, and contact with wildlife, performed by animal health officials' visits to farms within the control area. Direct contact with three hatcheries was identified during the traceback. One of the hatcheries was located in Rio Grande do Sul and contained 1,905,156 fertilized eggs and 68,000 1-day-old chicks. The other two were located outside the state, in Minas Gerais, with ~7 million eggs and ~11 million eggs on the other farm in Paraná. In Minas Gerais, the cleaning and egg removal commenced on May 16 (Day 6) and concluded on May 19 (Day 9), with 7,389,249 eggs being destroyed and 300,000 1-day-old chicks. On May 18 (Day 8), seven road barriers were installed to control vehicle access into the control zones until June 17 (Day 38).

The first round of clinical inspections of susceptible farms within the established control area around the infected farm was completed by May 20th (Day 10); a total of 2,113 farms were inspected and did not present any clinical signs of HPAI. Between May 22 (Day 12) and May 28 (Day 18), follow-up inspections in the control area started and were completed on May 31 (Day 21). From Day 21 on, the surveillance continued at the index farm, and monitoring was also conducted in high-risk municipalities with the detected presence of migratory bird routes throughout the state. The 28-day isolation and quarantine period ended on June 18, 2025 (Day 39), and business continuity was restored for the entire state. As of the writing of this manuscript, no new HPAI outbreaks have been detected in commercial poultry facilities in Rio Grande do Sul nor in Brazil. In farms within a 10-kilometer radius of the outbreak, as well as at farm supply stores and nearby schools, educational activities were conducted both during and after the quarantine period. The concurrent outbreaks of wild birds prompted wildlife authorities to intensify surveillance for sick or dead wild birds in the control areas. No additional mortality has been reported at the zoo since June 24, 2025, which led to the end of the quarantine for the zoo on July 8, 2025. The facility reopened to the public on July 31, 2025. A detailed description of each activity is presented in Supplementary Material Table S1, and the timeline of the outbreak is represented in Figure 3.

### *3.2 Forecasting the outbreak dynamics*

The MHASpread transmission model was used to assess the probability of secondary poultry infection from the infected farm. Our analysis revealed that, if HPAI was detected 3 days after its introduction into the commercial poultry breeding farm, the predicted median number of secondary infected farms was 4 (interquartile range [IQR]: 2–5; maximum: 12) (Figure 4). In the scenario where we assume the introduction occurred 5 days after the introduction, the model estimated a median of 6 (IQR: 3–22; maximum: 55) secondary infected farms. Notably, in scenarios where control actions were delayed for 10 days, the median number of infected farms reached 34 (IQR: 12–47; maximum: 59).

### 3.2.1 Spatial probability of infection

The MHASpread model was used to generate maps illustrating the regional probabilities of infection for the scenarios in which it was assumed that five days had passed since the disease was introduced (Figure 5). The municipality of Montenegro, the epicenter of the outbreak, exhibited the highest risk of infection. Interestingly, the model also highlighted two additional high-risk regions situated to the north of the primary outbreak area. Additional maps detailing the scenarios with 3- and 10-day delay detections are provided in the Supplementary Material, Figures S3 and S4.

### 3.3 Spatial risk analysis

To assess the spatial risk of HPAI in Rio Grande do Sul state, we integrated environmental factors with wild bird migratory flyway areas (Supplementary Material Figure S5) and poultry farm density with the spatial distribution of HPAI outbreaks (Supplementary Material Figures S6-S7 and Supplementary Material Table S3). The spatial regression models had the best fit to the data (Supplementary Material Table S4). The final spatial regression model selection retained temperature and water occurrence as predictor variables (Supplementary Material Table S5). Results were aggregated at the municipality level. Most municipalities in Rio Grande do Sul (71.23%) exhibited a very low risk of HPAI introduction, with values ranging from 0 to 0.001. In 11.87% of the municipalities, the risk was between 0.001 and 0.003, while 6.24% were in the range of 0.003 to 0.005. A limited proportion of municipalities showed higher risk values: 3.02% fell within the range of 0.005 to 0.009, and 1.81% within the range of 0.009 to 0.015. 5.83% (29 municipalities) exhibited the highest estimated risk, from 0.015 to 0.229 (Figure 6).

### *3.4 Outbreak response and control measures*

The control area was divided into two zones: an infected zone, comprising a 3 km radius around the outbreak, and a surveillance zone, comprising a 10 km radius around the outbreak. The 55 sites visited in the infected zone were all backyard farms, whereas among the 631 farms in the surveillance zone, two were commercial breeding farms (Figure 7). Farms in the infected zone were visited every three days, while those in the surveillance zone were visited every seven days (Supplemental Material Table S1). A total of 2,113 surveillance visits were conducted, during which seven suspected sites were identified; however, all samples taken were negative for HPAI. The most frequently observed domestic bird species of the 716 visited sites was the chicken (*Gallus gallus domesticus*). Ducks (*Anas platyrhynchos domesticus*) were observed at 164 locations, followed by guinea fowl (*Numida meleagris*) in 90 locations, domestic geese (*Anser anser domesticus*) in 80 locations, 76 locations with Turkeys (*Meleagris gallopavo*), while other species were recorded in 86 locations.

*3.4.1 Movement standstill, and road barriers*

A movement standstill was implemented for all bird movements into and out of the control area from May 15 (Day 5) to May 30, 2025, while movements outside the control area remained unrestricted. The restriction was lifted 13 days later. Six fixed road sanitary barriers were installed, and one mobile road barrier was utilized to control movements into and out of the control area (Figure 7). Barrier checkpoints resulted in 4,197 vehicle interceptions and the disinfection of a daily median of 131 (IQR: 60-243; maximum of 772, totaling 2,674 vehicles) (Supplementary Material Figure S9). The intercepted movements were classified as follows: 45 food supply trucks, 92 milk tank trucks, 175 vehicles transporting live animals, 715 passenger vehicles, and 3,161 vehicles for other purposes, such as those transporting meat products, eggs, and fruit.

*3.4.2 Staff and personnel*

The personnel deployed to work on the outbreak response included a wide range of state employees, such as veterinarians, managers, technical advisers, and support staff. These personnel were responsible for executing surveillance and control measures, thereby managing the primary field activities of the response. Federal animal health officials (federal staff): Officials and specialists from the Federal Ministry of Agriculture, Livestock and Supply provided national-level coordination, laboratory confirmation and sequencing, and policy guidance, and served as the interface with international institutions such as the WOAH. Local government employees from the Montenegro municipality were involved in on-the-ground coordination, collaborating with state animal health officials. Cleaning and disinfection of the infected farm were performed by the poultry integrator's staff under the supervision of official veterinary services. Military police security forces were deployed to secure the control area, enforce movement restrictions, secure road barriers, and ensure safe passage for emergency and official vehicles. Fire department emergency-response teams assisted with disinfection operations, disposal of culled animals and waste materials, decontamination of equipment, and ensuring biosecure conditions at disposal and burial sites. Other roles were filled by personnel who played various roles, such as administrative assistants, data entry personnel, laboratory support staff, communications officers, and logisticians, all of whom contributed to the multifaceted response effort.

The median daily number of state staff involved in the outbreak response was 24 (IQR: 0-54, maximum = 90), followed by military police with a median of 6 (IQR: 0-30, maximum = 50) (Figure 8). Other roles, including federal staff, the Montenegro municipality, the fire department, and other personnel, ranged from zero to six people.

## 4. Discussion

This manuscript describes the first HPAI outbreak in a commercial poultry farm in Brazil. The outbreak affected a breeder layer facility in Montenegro municipality, Rio Grande do Sul state, with a capacity of 19,040 birds. The infected farm exhibited 92% mortality before control actions were implemented. Sudden mortality was reported on the night of May 11, 2025. The farm was visited the next day, and samples were collected and submitted for testing. HPAI clade 2.3.4.4b was confirmed on the fourth day post notification to the state animal health officials. Immediate control actions included quarantine, culling, surveillance, and field investigations within the control area. From May 17th to 30th (fourth day to 13th post-notification), a total of 4,197 vehicles entering and exiting the control area were inspected at road disinfection barriers. While most birds died of HPAI, the remaining birds were depopulated, discarded and buried within the farm premises. Given the similarity of the HPAI strain circulating in South America and the detection in the zoo located 26 km from the infected farm, was similar to the strain in the infected farm, the most likely introduction could be attributed to wild birds. Biosecurity breaches, a lack of cleaning and disinfection of shared vehicles, equipment, and fomites, personnel crossing the farm perimeter buffer area, or issues with the fence used as the perimeter buffer area, could not be ruled out. As part of the contingency strategy during the outbreak response, a disease spread model was used to estimate the probability of secondary infections.

The most likely route of HPAI introduction in Brazil and its subsequent spread to commercial farms could be attributed to migratory waterfowl. Evidence suggests an established presence of HPAI in the wild bird population of the wider region, as the municipality of Montenegro lies in a river valley with rice fields and wetlands, key stopover sites for migratory birds (Zuquim Antas, 1994). In addition, the state of Rio Grande do Sul is located along both the Atlantic Americas and the South Atlantic flyway routes, which bring migratory birds from North America, Europe, and sub-Antarctic regions to southern Brazil during the winter (Graves, 1992; Zuquim Antas, 1994).

Laboratory results identified that the breeding farm was infected with HPAI clade 2.3.4.4b. This H5N1 strain is maintained in waterfowl, shorebirds, and seabirds(Reischak et al., 2023). Also closely related to strains circulating in Chile and Peru (Reischak et al., 2023) and in the Brazilian wild bird population (De Araújo et al., 2024). As of June 2025, no farm-to-farm transmission has been registered in Rio Grande do Sul, suggesting a single-point introduction. In addition, this entry route hypothesis is supported by concurrent HPAI detections in wild birds, including outbreaks among Black Swans and Black-necked Swans at the nearby zoo, which exhibit genetic nucleotide similarity values ranging from 98.21 % to 99.79 % with the commercial infected farm. Thus, the simultaneous detection of HPAI in zoo birds further suggests that wild birds, rather than backyard farms, are the most likely source of HPAI to the first commercial poultry farm infected in Brazil (Araujo et al., 2018). Moreover, our spatial analysis suggested that the distribution of migratory birds has been associated with the increase of HPAI outbreaks in Brazil, a finding consistent with previous research conducted in other regions (McDuie et al., 2022; Ramey et al., 2022).

      The first HPAI infection in a commercial poultry farm in Brazil marked a turning point in Brazil's HPAI epidemiology, as the virus breached the high biosecurity premises of a commercial breeding poultry operation. The spread of HPAI from wild to backyard birds and now into commercial poultry underscores the growing threat of wild bird-mediated transmission, particularly in regions with extensive wetlands and high migratory bird activity (Araujo et al., 2018). The rapid response, including laboratory diagnosis confirmed within four days of notification and sampling, was critical in containing the outbreak within 39 days. In comparison, similar outbreak durations were observed in other countries with a comparable number of infected farms. For instance, Japan successfully stamped out an outbreak in 49 days through culling, movement controls on poultry, and intensive surveillance (Sugiura et al., 2006). Another important point is that historical HPAI control has focused on culling and quarantine, which successfully eliminated the 1959 Scottish outbreak and the 1997 Hong Kong H5N1 outbreak. Both outbreaks were effectively controlled by depopulating the entire domestic poultry population within the affected zones (Alexander and Brown, 2009). In European countries, the use of quarantine, movement controls, and mass culling proved highly effective in controlling early HPAI outbreaks, and in the late 19th century and in North America (Alexander and Brown, 2009; Kirkeby and Ward, 2022). However, not all outbreaks have been eradicated quickly. More recently, the ongoing HPAI H5N1 panzootic in the U.S., which has been active since 2021,

presents an unprecedented challenge. This outbreak has affected over 134.72 million birds in the U.S. since February 2022, resulting in significant economic losses estimated at over USD 1 billion (Animal and Plant Health Inspection Service, 2024). This prolonged and costly crisis raises serious questions about the long-term sustainability of relying solely on the culling policy (Swayne, 2012). Globally, under such conditions, some countries are evaluating the use of vaccination in poultry as a control tool, where systematic vaccination serves as an additional layer of prevention, to be used in conjunction with traditional measures, as implemented in a region of France (Degeling et al., 2016; Planchand et al., 2025). Given the successful elimination of the first Brazilian HPAI outbreak, the use of preventive or emergency vaccination has not yet been considered. However, we remark the need for discussions on whether emergency use of vaccines could be part of the contingency toolbox if HPAI becomes seasonal or endemic, as in North America.

**Limitations and further remarks**

Several limitations have been identified during the response to the outbreak. First, as filed investigations began within the control area, the geolocation of all premises or properties with birds was visited and geolocated. Some of these geolocations were visited more than once, per the number of farm visit rounds shown in Figure 1 and Supplementary Material Table S1. Thus, when we do not present the number of backyard farms in the control area, alternatively, we show the number of geolocations visited. Second, we have used HPAI transmission parameters derived from studies conducted outside Brazil (Antonopoulos et al., 2024; Comin et al., 2011; Nickbakhsh et al., 2016), since Brazil-specific transmission data for HPAI were not available at the time we modeled the HPAI dynamics. While these parameters provide the best approximation under current circumstances, future modeling efforts should prioritize locally generated data to better capture the dynamics observed in Rio Grande do Sul. Finally, we emphasize the need to further develop our transmission model, which should include additional transmission pathways, such as the transmission from and to wild birds, and also incorporate egg movements as an additional pathway of transmission.

**Conclusion**

The first HPAI outbreak in a commercial poultry farm in Brazil highlights the importance of the state's animal health services' level of preparedness and response, which resulted in the complete elimination in 39 days. The modeling indicated that if HPAI was indeed detected within the three days, as indicated based on the outbreak investigation, the epidemic would be restricted to a few farms. It turns out that due to the rapid implementation of control measures, the epidemic was stamped out without any secondary infections. Given that the outbreak occurred in a commercial breeder facility with a high level of biosecurity, it was insufficient to prevent the introduction of HPAI. While the main introduction route could not be confirmed, it was likely driven by wild birds and/or biosecurity breaches. The successful use of an informatized system (PDSA) further demonstrates the value of integrated information systems in coordinating real-time outbreak responses. Moreover, the mobilization of diverse stakeholders, including federal and state agencies, the military, and local authorities, underscores the importance of intersectoral coordination and a high volume of human resources in addressing animal health emergencies.

**Data availability statement**

A series of official reports titled COEZOO-RS is available on the official Rio Grande do Sul state website (https://www.agricultura.rs.gov.br/inicial). The data related to geolocation animal movements and farm identification are not publicly available and are protected by confidential agreements. Therefore, they are not available.

**Funding**

Fundo de Desenvolvimento e Defesa Sanitária Animal (FUNDESA-RS) award numbers 2021–1318 and 060496. Foundation for Research of the State of Rio Grande do Sul (FAPERGS) award number 24/2551-0001401-2. Vinícius Maran is partially supported by the Brazilian National Council for Scientific and Technological Development (CNPq), award number 306356/2020-1 (DT2).

**Figure legend**

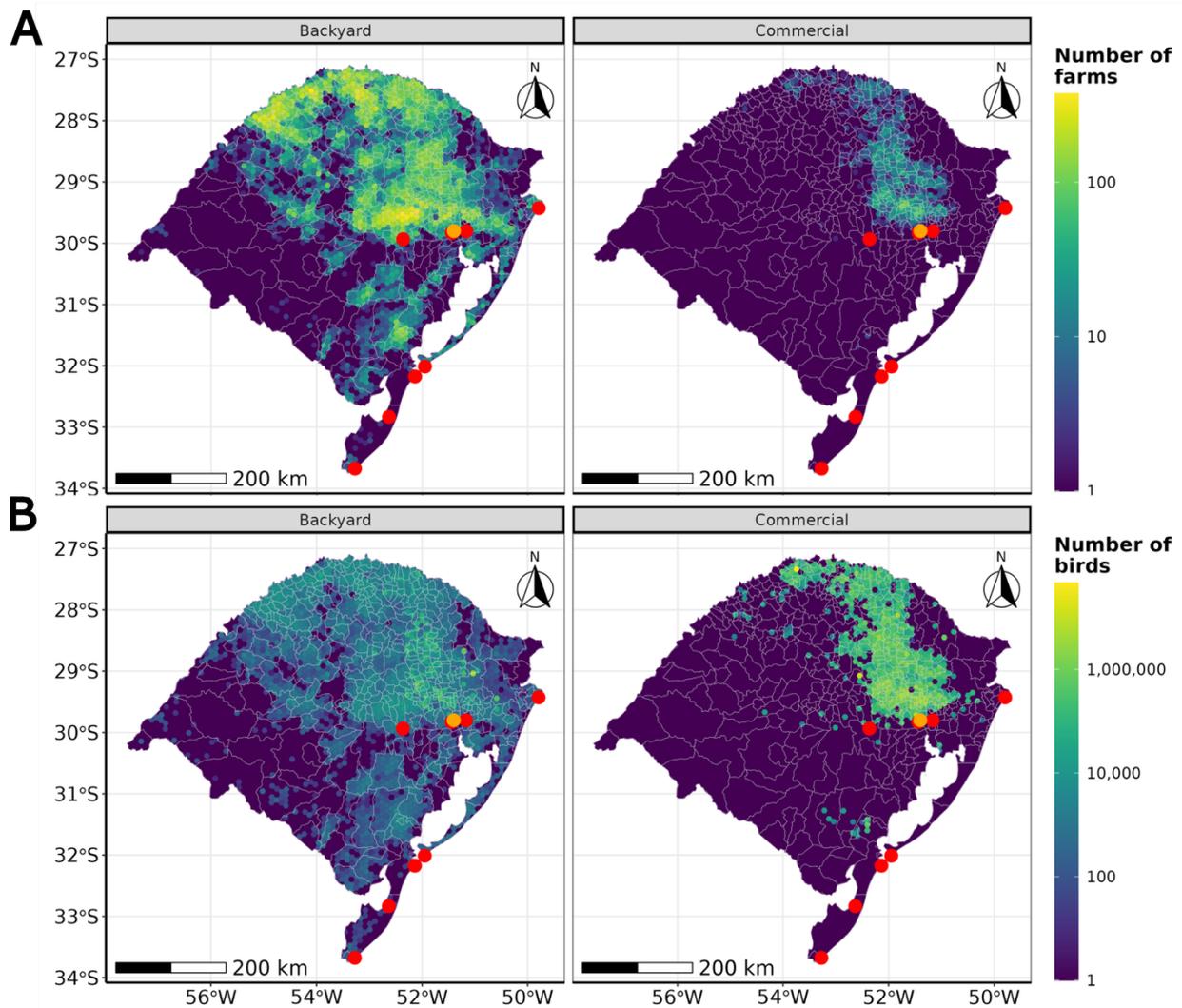

**Figure 1. Spatial distribution of poultry farms in Rio Grande do Sul, Brazil**. A) Farm density. Each hexagon grid is 10 km in size on a log10 scale. The hexagon color represents the total number of farms; B) Bird density. The hexagon color represents the total number of bids in each grid on a log10 scale. White lines represent the municipality's division. Red dots represent HPAI cases in wildlife reported since May 2023, including the following species: *Furnarius rufus, Arctocephalus australis, Otaria flavescens, Plegadis chihi, Cygnus melancoryphus,* and

*Thalasseus maximus*. The orange dot represents the outbreak in the HPAI-positive commercial breeding farm in the Montenegro municipality, Rio Grande do Sul, Brazil.

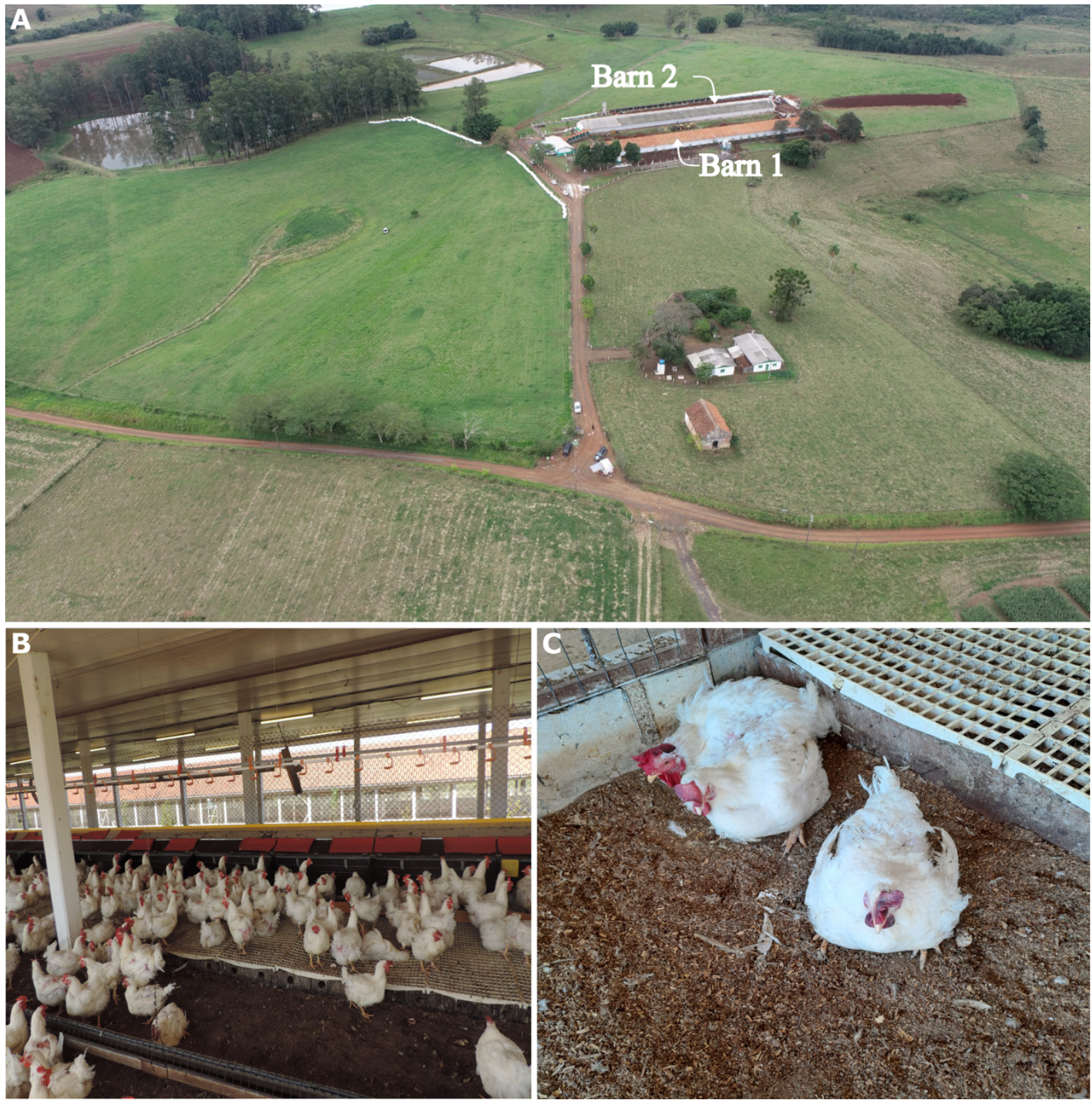

**Figure 2. Overview of the infected farm, housing conditions, and infected birds' clinical signs.** A) Aerial view of the infected farm showing the two affected barns and a perspective of the premises and biosecurity features (e.g., the farm was fenced). In the top right of the image,

the area was used for carcass and egg burial. B) View of barn two of the affected farm. C) Close-up of two infected breeder hens showing apathy, cyanosis, and prostration.

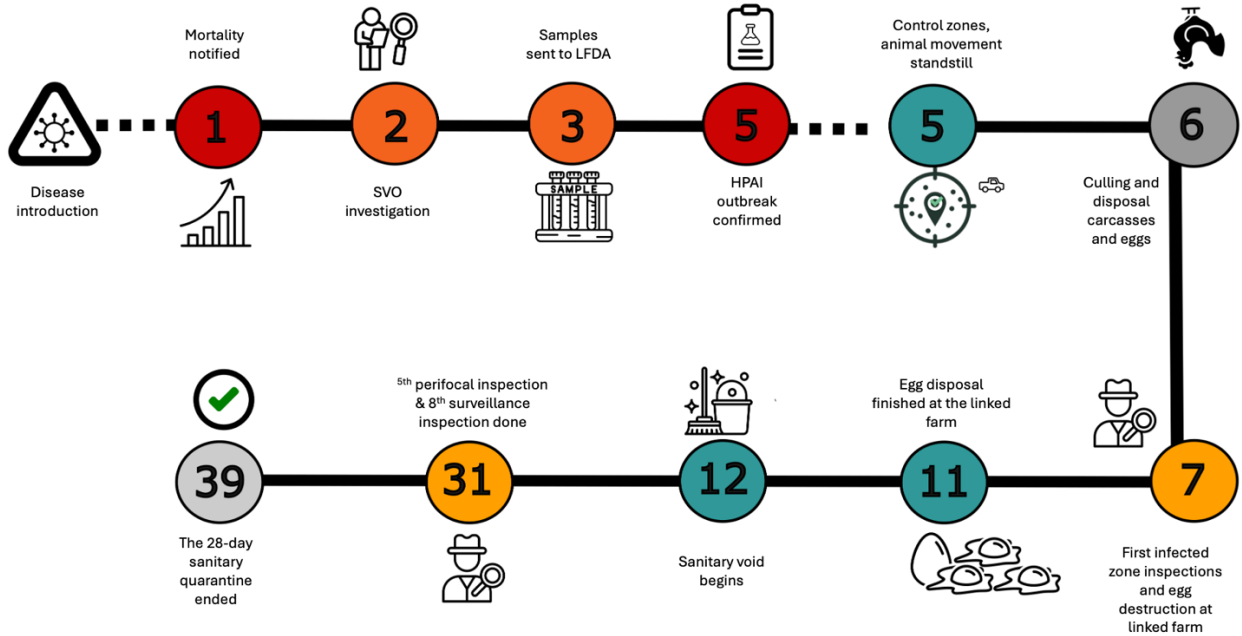

**Figure 3. Summary of the main events in the first HPAI outbreak in Rio Grande do Sul, Brazil.** The numbers inside the circles represent the actions taken on a specific day since the start of the emergency response. Each circled number corresponds to a milestone.

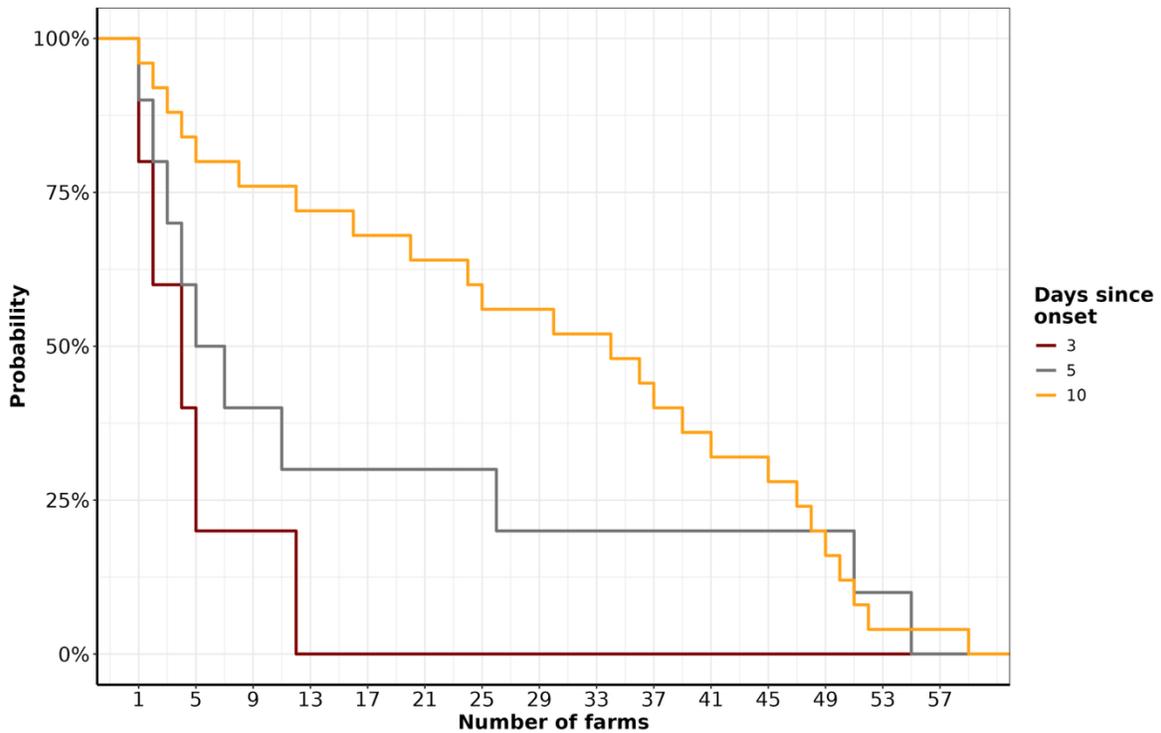

**Figure 4. Empirical cumulative distribution illustrating the simulated number of predicted secondary infections.** The y-axis shows the probability of new infected farms, and the x-axis indicates the number of infected farms. Each line represents the probability distribution of scenarios in which HPAI was detected within 3 days (red), 5 days (gray), and 10 days (orange), from when it was reported to the state animal health officials.

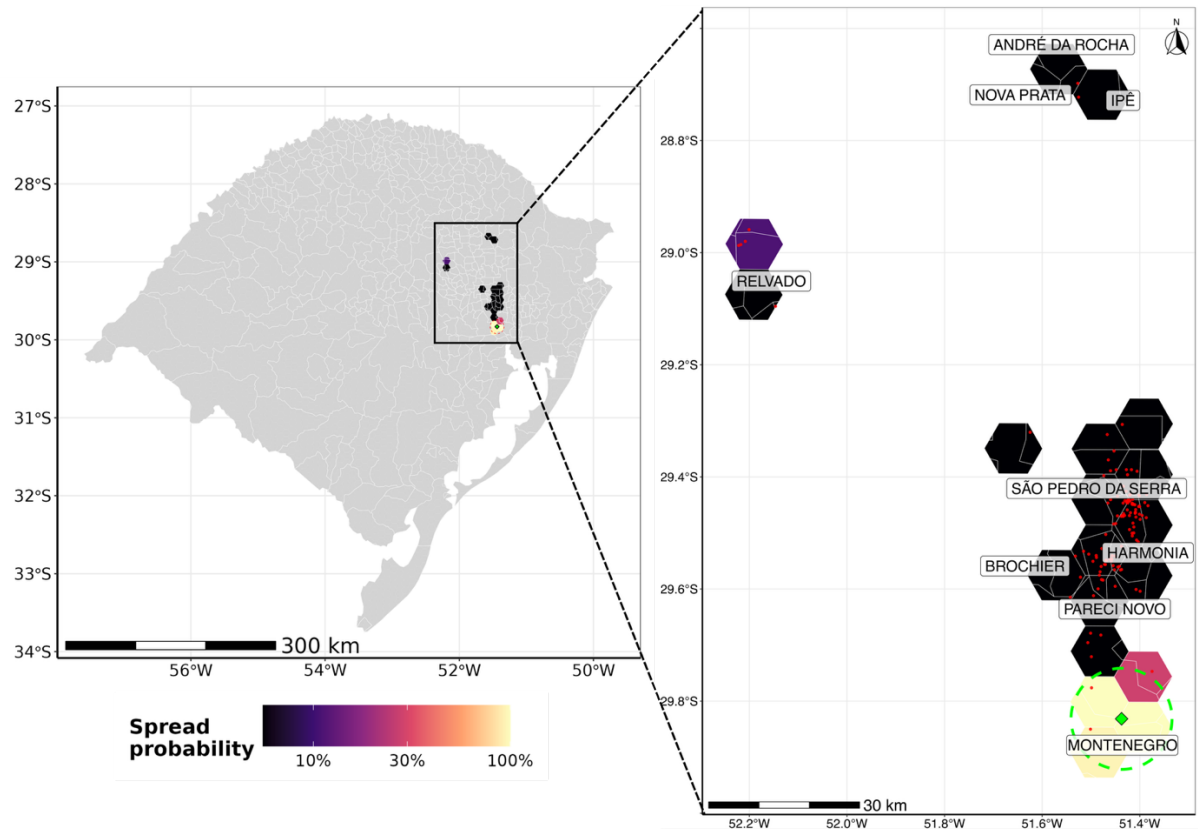

**Figure 5. Spatial probability of infection**. Results from the MHASpread model showing HPAI spread after five days from introduction without control measures. The left map provides an overview of Rio Grande do Sul, while the right panel provides a closer view of the area. The color scale quantifies the probability of finding at least one infected farm per 10 km hexagonal grid cell. Green dots represent infected farms, and red dots denote the geolocations colored with their probability of being infected.

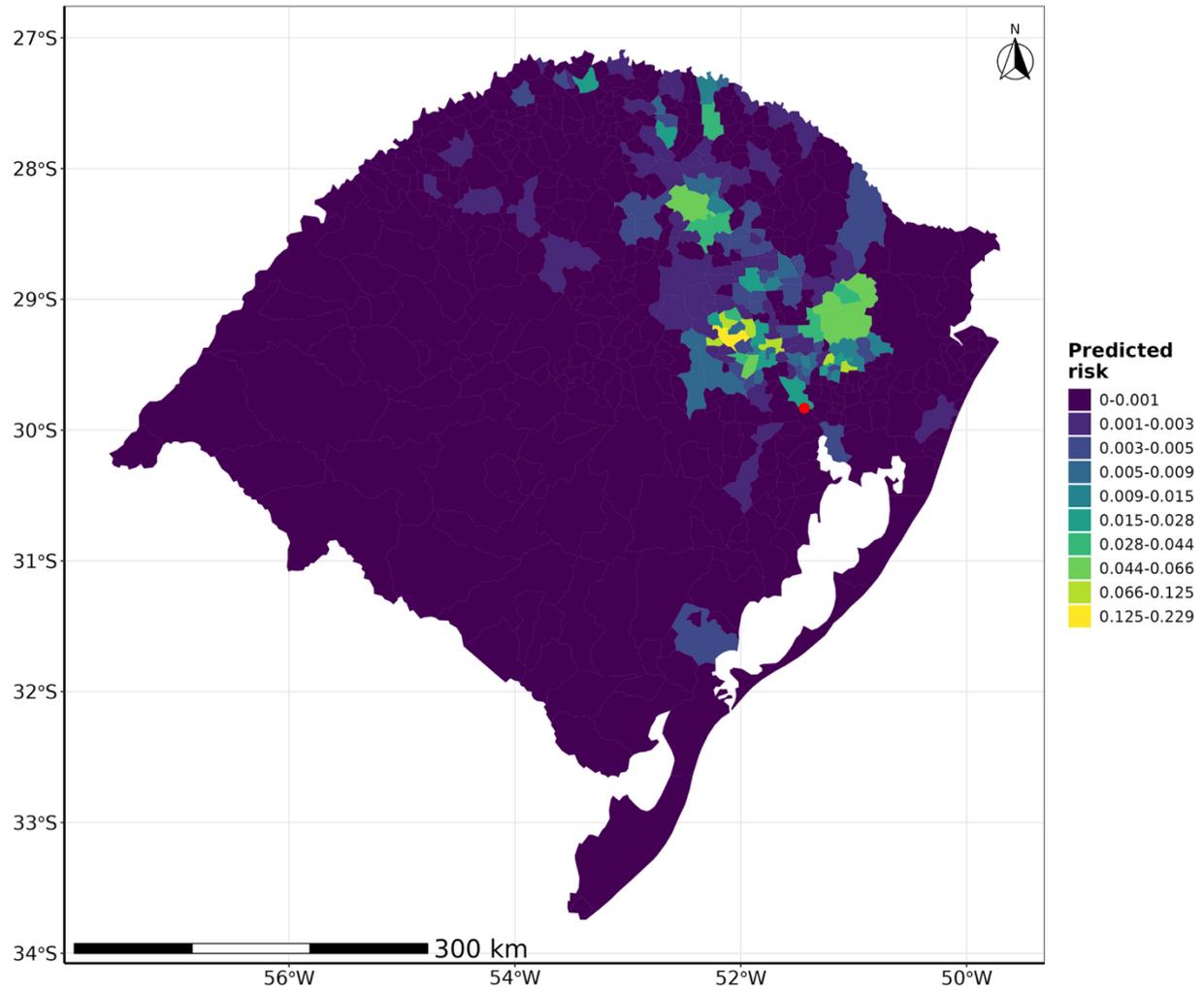

**Figure 6. Predicted spatial risk of HPAI introduction into commercial poultry farms by the municipality.** The color represents the risk value probability grouped by municipality level. The red dot represents the location of the outbreak in Montenegro in Rio Grande do Sul, Brazil.

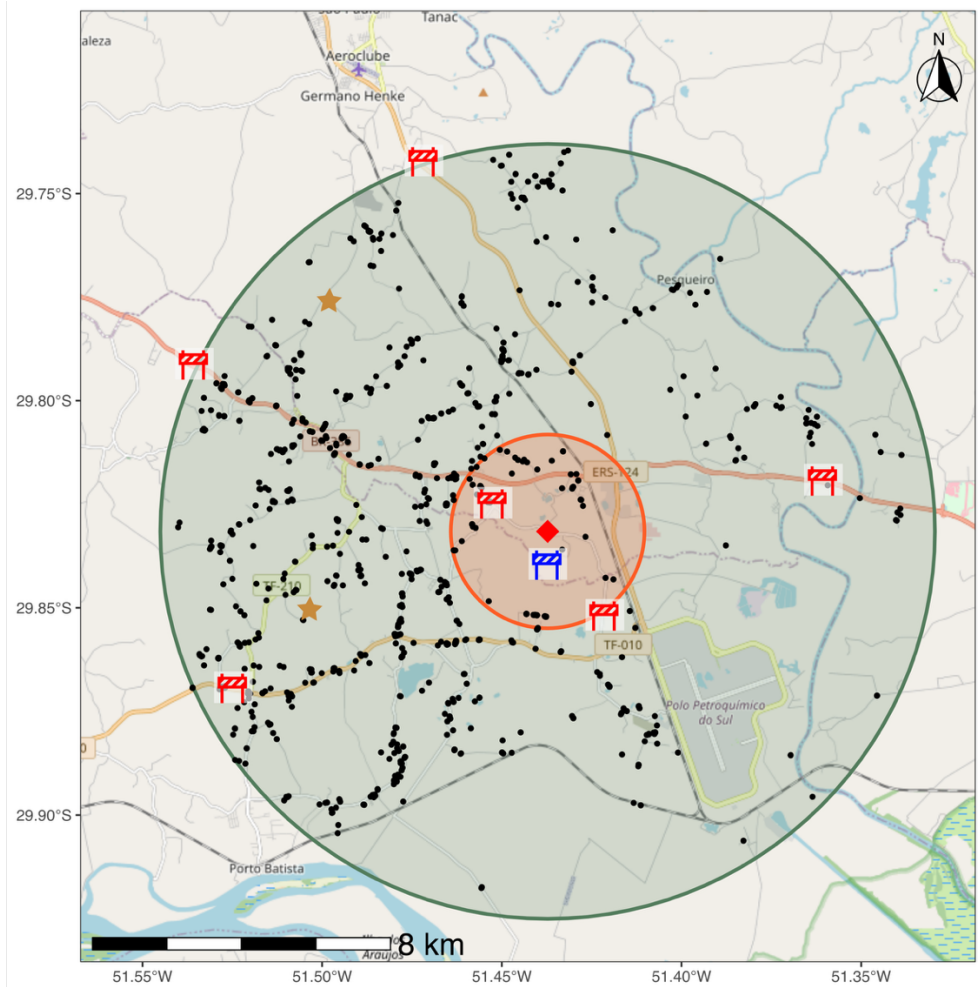

**Figure 7. Control area.** The central red dot represents the infected commercial breeding layer farm. In contrast, the black dots indicate the geolocations in which backyard poultry farms were located, and the brown star dots represent commercial breeding poultry farms. The orange inner circle represents the infected zone (3 km), with the green circle serving as a buffer zone (10 Km). The blue barrier icon indicates one of the locations of the mobile road barriers, while fixed road barriers are shown in red.

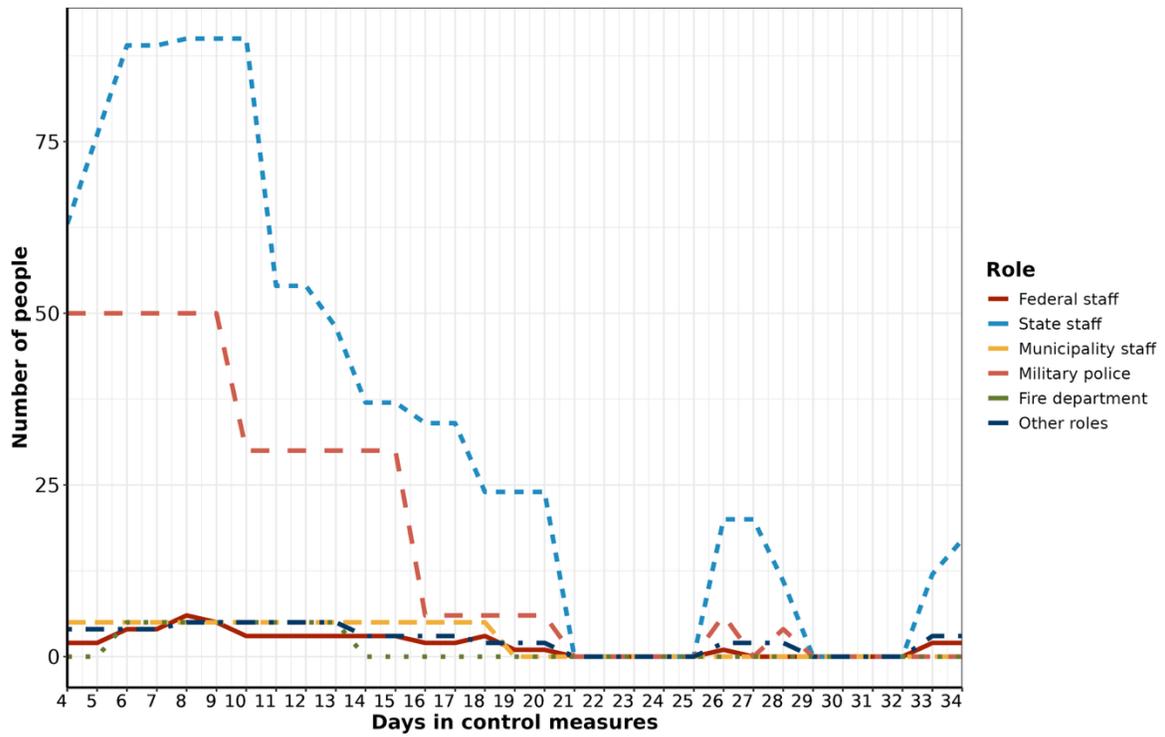

**Figure 8. Distribution of the number of staff involved in the first HPAI outbreak in Brazil.**
Daily count of personnel involved in control measures, by role. The time series shows the number of people (y-axis) engaged each day throughout the response (x-axis).